\documentclass{appolb}
\usepackage{amsmath,amssymb,multirow,epsfig,bm,color,graphicx}

\newcommand{\beq}{\begin{equation}}
\newcommand{\eeq}{\end{equation}}
\newcommand{\bea}{\begin{eqnarray}}
\newcommand{\eea}{\end{eqnarray}}

\newcommand{\fig}[1]{Fig.~\ref{#1}}

\newcommand{\eF}{\varepsilon_{F}}

\newcommand{\kF}{k_{\textrm{F}}}

\begin{document}
\title{Properties of spin-polarized impurities - ferrons, in the unitary Fermi gas%
\thanks{Presented at the XXXVI Mazurian Lakes Conference on Physics, Piaski, Poland, September
1-7, 2019.}%
}
\author{B. T\"uzemen$^{a}$, P. Kukli\'nski$^{a}$, 
P. Magierski$^{a,b}$, G. Wlaz{\l}owski$^{a,b}$
\address{$^{a}$Faculty of Physics, Warsaw University of Technology, ulica Koszykowa 75, 00-662 Warsaw, POLAND \\ 
$^{b}$Department of Physics, University of Washington, Seattle, WA 98195–1560, USA}
}




\maketitle

\begin{abstract}
A new excitation mode has been predicted to exist in the unitary Fermi gas. 
It has a form of a spin-polarized impurity, which was  
dubbed as ferron. It is characterized by a closed nodal surface of the pairing field
surrounding a partially spin-polarized superfluid region, where the phase differs 
by $\pi$. In this paper, we discuss the effect of temperature on the generation of the ferron and 
the adiabaticity of the spin-polarizing potential together with ferron's ground state properties.
\end{abstract}
  
\section{Introduction}
Pioneered by the works of Eagles~\cite{eagles} and Leggett~\cite{leggett} together with Nozi\`{e}res and Schmitt-Rink's extension to finite temperatures in 3D~\cite{nozieres}, the unitary Fermi gas (UFG) has been under the attention of ultracold communities. One of its remarkable features is a pairing gap almost as large as half the Fermi energy~\cite{bloch}. This exceptionally strong pairing field results in a coherence length which is of the order of interparticle distance. Therefore in UFG there is no clear separation 
of scales related to superconductivity and to single-particle motion.
The strong pairing in UFG admits also the possibility of having relatively large spin-imbalance without a loss of superfluid
properties. It offers the possibility to investigate superfluidity in the broken time-reversal symmetry framework, which
lead to consideration of exotic phases such as FFLO phase~\cite{ff,lo}, and Sarma phase~\cite{sarma}.

Recently we have reported the existence of a stable, spin-polarized impurity in ultracold Fermi gases~\cite{ferron}. This droplet, dubbed as {\it ferron}, consists of an excess spin population that does not have the necessary partners to form Cooper pairs. The procedure to dynamically generate the ferron involves a time-dependent, local and spin-selective potential. During its application to the uniform, unpolarized system, the potential repels one spin component and attracts the other thus creating a local spin polarization $p(\bm{r})=\frac{n_{\uparrow}(\bm{r})-n_{\downarrow}(\bm{r})}{n_{\uparrow}(\bm{r})+n_{\downarrow}(\bm{r})}$ by breaking the Cooper pairs. Introducing the local polarization to the system shifts the Fermi surface of spin-up and spin-down particles with respect to their new chemical potentials. Having two different Fermi surfaces causes oscillations in the phase of the pairing field~\cite{buzdin,Halterman} creating two regions for the pairing with a phase difference of $\pi$. This phase shift creates a nodal surface where the strength of the pairing field goes to zero. Even after the time-dependent potential is turned off once generated nodal surface harbors the polarized population~\cite{SuppressedSolitonicCascade}, creating the stable droplet, ferron. 

\section{Stability of the ferron}

The main mechanism behind the stability of the ferron lies in its resemblance to the superconductor-ferromagnet-superconductor (SFS) junctions. The nodal surface where the pairing drops down to zero acts as a normal-metal neighboring two superconductors causing Andreev states to emerge because of the proximity effect~\cite{buzdin}. The unpaired population occupies these Andreev levels, stabilizing the nodal surface and ensuring the impurity's exceptionally long life-time. 
The detailed discussion of ferron's dynamical generation and its stability can be found at Ref.~\cite{ferron}. 

\begin{figure}[hb]
\centerline{%
\includegraphics[width=0.8\columnwidth]{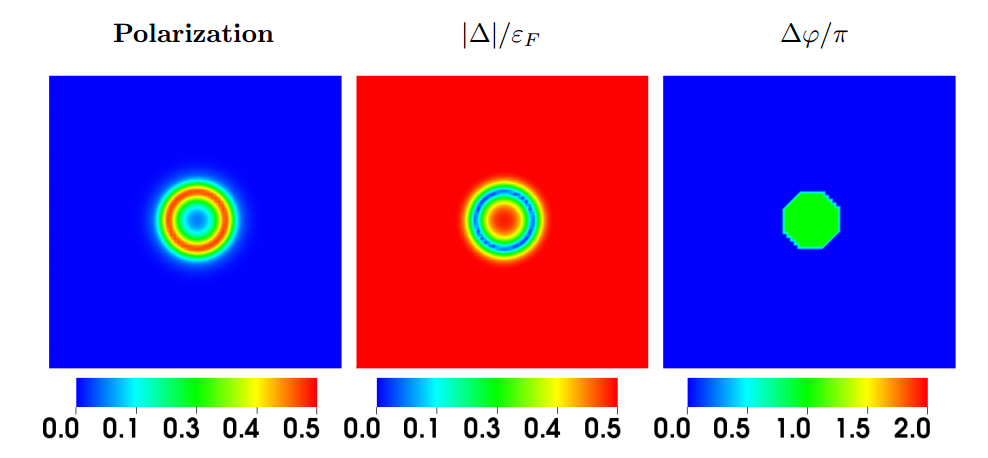}}
\caption{Ferron at its ground state. The local polarization $p(\bm{r})$ is shown on the left picture. The pairing strength $|\Delta(\bm{r})|$ and phase difference $\Delta\varphi(\bm{r})$ (measured with respect to the value far away from the ferron) are shown at the middle and right pictures respectively.}
\label{ferron_gs}
\end{figure}

In this paper, along with its dynamics, we focus on ferron's static properties. To investigate ferron, we employ superfluid local density approximation (SLDA) which is a framework originated from density functional theory extended to superfluid systems~\cite{LNP__2012}. While extending the density functional theory into Fermi superfluids, a great deal of difficulty arises from pairing correlations $\Delta(\boldmath{r},\boldmath{r^\prime})$. They are in principle non-local and give rise to a set of integro-differential equations. One way to overcome this problem is to introduce instead a local pairing field, $\Delta(\boldmath{r})$. In order to explore spin-imbalanced systems, we use asymmetric-SLDA (ASLDA) described in Ref. \cite{LNP__2012}, which has been tested against QMC calculations. The set of equations used in ASLDA is similar to well-known Bogoliubov de-Gennes equations. The equations have the following form:
\begin{equation}
\left( \begin{array}{cc}
 h_{\uparrow}(\textbf{r}) & \Delta(\textbf{r}) \\
 \Delta^*(\textbf{r}) & -h_{\downarrow}^*(\textbf{r})
\end{array} \right)\left( \begin{array}{cc}
u_{n,\uparrow}(\textbf{r}) \\
v_{n,\downarrow}(\textbf{r})
\end{array} \right) = E_n \left( \begin{array}{cc}
u_{n,\uparrow}(\textbf{r}) \\
v_{n,\downarrow}(\textbf{r})
\end{array} \right).
\label{hamiltonian}
\end{equation}
Here $h_{i}(\textbf{r})$ denotes the single-particle Hamiltonian. It consists of kinetic, mean-field, and chemical potential terms. The solutions for the spin reversed components of quasi-particle wavefunctions can be obtained via the symmetry relation $u_{n,\uparrow}\rightarrow v_{n,\uparrow}^*$,  $v_{n,\downarrow}\rightarrow u_{n,\downarrow}^*$ and $E_n \rightarrow -E_n$ where $E_n$ is the quasiparticle energy. The detailed explanation of the ASLDA and its extension to time-dependent problems can be found at Ref.~\cite{LNP__2012}.

In dynamical calculations of Ref.~\cite{ferron}, we start with an unpolarized, uniform UFG. This is followed by the application of spin-selective, time-dependent potential giving rise to the nodal surface by breaking the Cooper pairs. Differently, in static calculations that we perform in the present work, in order to obtain the ground state solution for the ferron, we start with a spin-imbalanced UFG. We imprint the phase difference by hand to create the nodal surface in a given radius. Hence, the polarized population can occupy the Andreev states localized within the nodal surface. The static solution is obtained
from the stationarity (minimum) condition of the quantity:
\begin{equation}
\left\langle\Omega\right\rangle = \left\langle H-\sum_{\sigma=\{\uparrow,\downarrow\}}\mu_{\sigma}N_{\sigma}\right\rangle 
\end{equation}
where $\mu$ and $N$ are chemical potentials and particle numbers, respectively. The lattice size in our static calculations is $80\kF^{-1}$ in x and y directions, and the BCS coherence length $\xi$ is about $1.27\kF^{-1}$ where $\kF=(6\pi^2 n_{\uparrow})^{1/3}$ is the Fermi momentum (we use metric system where $m=\hbar=k_B=1$). In z-direction, we expand the wavefunction in plane waves, therefore, a circular phase difference in 2D yields a tube-shaped ferron in 3D. The structure of ferron in its ground state, including its local spin polarization and pairing field distributions is shown in~\fig{ferron_gs}.

The ferron radius is clearly dependent on the total spin-imbalance, as the spin polarization is associated
with occupation of Andreev states localized around the pairing nodal surface. On the other hand the number of Andreev states
depends on the ferron volume. In order to investigate this effect quantitatively 
we considered a UFG with a given total polarization $P=\frac{N_{\uparrow}-N_{\downarrow}}{N_{\uparrow}+N_{\downarrow}}$. Initially, we applied a phase imprint procedure in two regions by setting the value of the pairing field in Eq. (\ref{hamiltonian}) in the following way:

\begin{equation}
\Delta(\bm{r})=\left\lbrace 
\begin{array}{ll}
-\Delta, & r<R_{1},\\
\phantom{-}\Delta, & R_{2}<r,
\end{array}
\right. 
\end{equation}

\begin{figure}[hbt]
\centerline{%
\includegraphics[width=0.83\columnwidth, trim=0 140 0 130 clip]{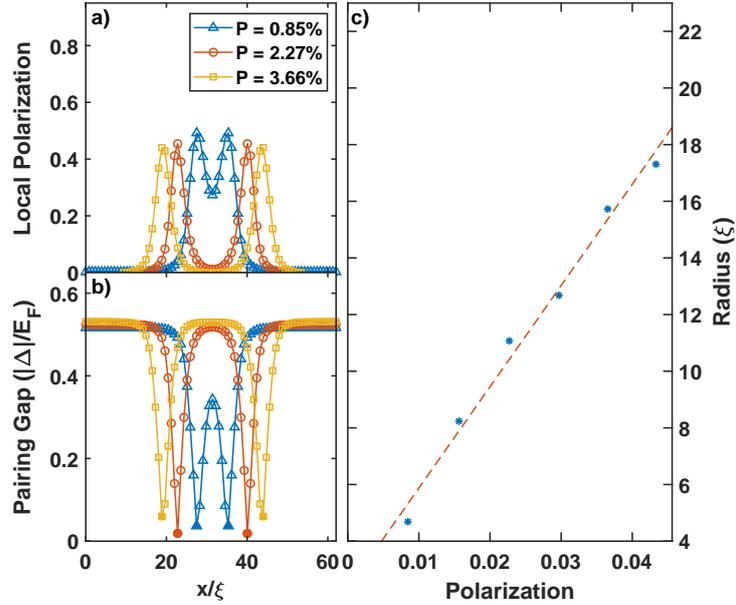}}
\caption{A profile along x-dimension of the system's local polarization (a) and the magnitude of the pairing field in the units of Fermi energy (b). The filled points on the lower values of panel b represent where the pairing field has its phase shifted by $\pi$. On the right panel (c) the radius of the nodal surface is given for different total polarizations. The dashed lines are a guide to the eye. The box size is 62.8$\xi$ on x and y directions. On z-direction the plane wave solutions are applied. Here, $\xi$ represents the coherence length.}  
\label{ferron_profile}
\end{figure}

This constraint allows to converge to the ferron-like structure.
After implementing the above condition for the desired number of iterations, the constraint is released and the system converges to its ground state with a radius between $R_1$ and $R_2$. Hence varying $P$ we can arrive at static solutions
corresponding to various ferron sizes as shown in Fig. \ref{ferron_profile}.

The radius of the ferron's lowest energy configuration is a function of the total polarization of UFG. Numerically we have found that there is a linear relation between the polarization and the radius. Since the unpaired particles occupy the Andreev states inside the nodal surface, the increase of polarization has to be accompanied by the increase of the number of Andreev states and thus require the ferron size to grow. In~\fig{ferron_profile} it can be seen that for small values of polarization the radius of the nodal surface is so small that the pairing gap is not able to completely recover itself thus allowing a non-zero polarization inside the ferron. 

\section{Ferron at finite temperature}

The stability of the ferron is governed by the interplay between the spin-imbalance and the pairing field. Clearly
the thermal excitations of UFG will weaken the stability of the ferron leading eventually
to its collapse. One may ask however a non obvious question, whether the instability occurs at the critical
temperature for normal-to-superfluid transition or for $T<T_{c}$. 

\begin{figure}[ht]
\centerline{%
\includegraphics[width=0.8\columnwidth, trim=0 160 0 160 clip]{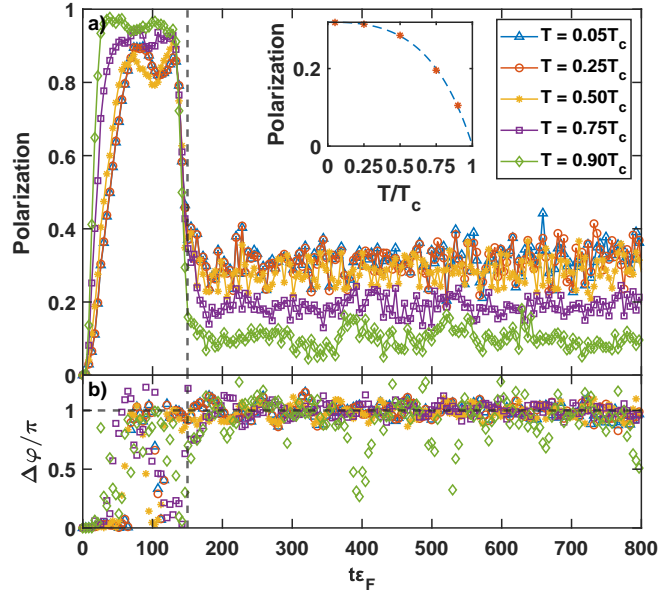}}
\caption{Effect of non-zero temperature on the ferron formation. Panel a) shows the time evolution of local polarization in the center of the object. The vertical dashed lines shows where the spin-polarizing potential is turned off ($t\eF=150$). Panel b) shows the pairing phase difference. The box size is $40^3$, corresponds to the length of $31\xi$ along 
each dimension. The amplitude of the potential was fixed at $A_0=2\eF$. The width of the spin polarizing potential is set to $\sigma = 4.71\xi$. (Inset) The time averaged local polarization in the center of ferron for different temperatures. The blue dashed lines are a guide to the eye to make the trend more visible.}
\label{fig_polar_temp}
\end{figure}

To check ferron's stability as a function of temperature, let us first perform dynamical calculations similar to Ref.~\cite{ferron}. We apply the ASLDA framework
described in Ref. \cite{LNP__2012}. Therefore the particle, kinetic, anomalous and current densities are 
calculated according to the assumption of having the equilibrium system represented by the grand canonical ensemble ($i=\{\uparrow,\downarrow\}$ labels spin components):
\begin{eqnarray}
n_{i}(\textbf{r}) &=& \sum_{|E_n|<E_c} |v_{n,i}(\textbf{r})|^2 f_{\beta}(-E_n), \\
\tau_{i}(\textbf{r}) &=& \sum_{|E_n|<E_c} |\nabla v_{n,i}(\textbf{r})|^2 f_{\beta}(-E_n), \\
\nu(\textbf{r}) &=& \sum_{|E_n|<E_c}  v_{n,\downarrow}^{*}(\textbf{r}) u_{n,\uparrow}(\textbf{r})\frac{f_{\beta}(-E_n)-f_{\beta}(E_n)}{2}, \\
\textbf{j}_{i}(\textbf{r}) &=& \sum_{|E_n|<E_c} \textrm{Im}[v_{n,i}(\textbf{r}) \nabla v_{n,i}^{*}(\textbf{r})]f_{\beta}(-E_n),
\end{eqnarray} 
where $E_n$ is the quasiparticle energy and $E_c$ is the cut-off energy as required by the regularization procedure. The densities are defined in terms of Bogoliubov quasi-particle wave functions $\{ v_{n,i}, u_{n,i} \}$. The Fermi-Dirac distribution, $f_{\beta}(E) = 1/(\exp(\beta E)+1)$, where $\beta=1/T$, allows us to model finite-temperature effects. In order to scale the temperature, we use the well-known BCS result that connects the pairing gap and the critical temperature, i.e. $\Delta/T_c = 1.76$~\cite{schrieffer}. Since the system under observation is in the unitary regime, it should be noted that the following results are only qualitative.

We start with a non-polarized, uniform UFG at finite $T$. Subsequently we apply a spin-polarizing, time-dependent Gaussian potential $V_{i}(\textbf{r},t) = \lambda_{i} A(t) e^{-r^2/2\sigma^2}$ of width $\sigma$ and amplitude $A_{0}=\max[A(t)]$. 
The potential repels spin-up atoms ($\lambda_{\uparrow}=1$) and attracts spin-down atoms ($\lambda_{\downarrow}=-1$). 
It is applied until the nodal surface appears as a result of breaking the Cooper pairs. Once the nodal surface is generated, the potential is removed from the system.
The described procedure simulate the creation of ferron at finite temperature, however it is not a fully consistent approach.
Namely, it is assumed that the densities evolve in time according to amplitudes $U$ and $V$, but their thermal distributions
are calculated in equilibrium, since the quasiparticle energies are fixed at initial values. One may expect that
the method may lead to reasonable results in the case of small departure from equilibrium and it may artificially enhance
the ferron stability. In~\fig{fig_polar_temp} we present the dynamical creation and stability of the ferron within our framework.

\begin{figure}[hbt]
\centerline{%
\includegraphics[width=0.8\columnwidth, trim=0 160 0 170 clip]{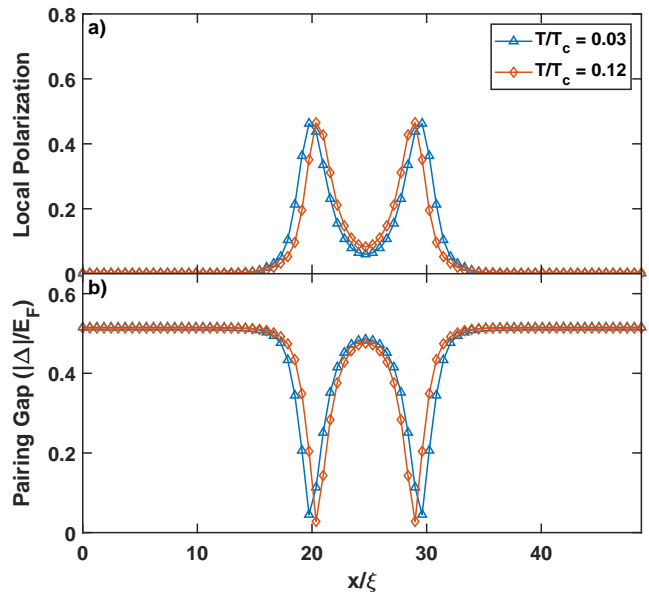}}
\caption{A profile along x-dimension of the system's local polarization (a) and the magnitude of the pairing field in the units of Fermi energy (b). The highest temperature in which the ferron can exist is shown in comparison with a colder system.}
\label{static_temp}
\end{figure}

Although dynamic calculations indicate that ferron structure is stable for $T<T_{c}$ it is instructive
to investigate whether stable, static ferron-like solution exists as well.
Therefore we perform a series of calculations at finite T according to the following procedure.
We first generated a ferron at $T = 0.03T_{c}$ by minimizing the density functional at fixed total spin-polarization. 
Subsequently the system is heated up gradually and at each $T$ it is checked whether
the ferron-like structure exist. For the polarization of $P = 1.5\%$, at $T > 0.12 T_{c}$ we observe that the ferron collapses due to the thermal excitations. During the gradual increase of the system's temperature, the ferron's radius slightly shrinks (see~\fig{static_temp}). This is accompanied with an increase in the local polarization. 
The obtained result that suggest that there is a critical temperature above which the static ferron-like solution is no longer stable.

\section{The role of adiabaticity in the ferron creation process}

In dynamical procedure to generate the ferron, we applied a spin-selective external potential in the form of:
\begin{equation}
V_{i}(\bm{r},t)=\lambda_{i} A(t) \exp\left[  -\frac{x^2 +y^2 +z^2}{2\sigma^2} \right]. 
\label{SMeq:V_p}
\end{equation}
Here, $\lambda_{i}$ stands for spin-up and spin-down particles and $\sigma$ sets the width of the Gaussian potential. The amplitude, $A(t)$ is a function of time, given by:
\begin{equation}
A(t)=\left\lbrace 
\begin{array}{ll}
 A_0\,s(t,t_{\textrm{on}}), & 0\leqslant t<t_{\textrm{on}},\\
 A_0, & t_{\textrm{on}} \leqslant t<t_{\textrm{hold}},\\
  A_0\,[1-s(t-t_{\textrm{hold}},t_{\textrm{off}}-t_{\textrm{hold}})], & t_{\textrm{hold}}\leqslant t<t_{\textrm{off}},\\
  0, & t\geqslant t_{\textrm{off}}.
\end{array}
\right. 
\label{SMeq:At}
\end{equation}
Here, $A_0\approx 2\eF$ denotes the amplitude of the potential where $\eF$ is the Fermi energy. The switch function, $s(t,w)$ is used to apply the potential in a desired rate and allows to control
adiabaticity of the process. It ranges from 0 to 1. The switching rate is defined by:
\begin{equation}
 s(t,w)=\dfrac{1}{2}+
 \dfrac{1}{2}\tanh\left[\tan\left(  \frac{\pi t}{w}-\frac{\pi}{2} \right) \right].
 \label{eq:switch}
\end{equation}

\begin{figure}[htb]
\centerline{%
\includegraphics[width=0.8\columnwidth, trim=0 160 0 160 clip]{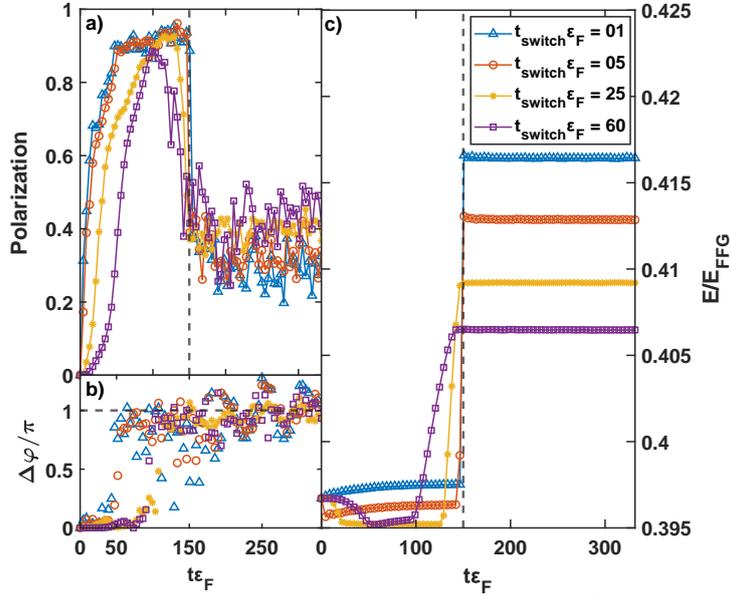}}
\caption{Time evolution of ferron under different switching-on/off rates of the generating potential. Panel a) shows the time evolution of the local polarization in the center of the object. Panel b) shows the pairing phase difference. In the panel c) the effect of different switching-on/off rates of the generating potential on the total energy of ferron is seen. The vertical dashed lines shows where the spin-polarizing potential is turned off ($t\eF=150$).}
\label{fig_etot_adia}
\end{figure}

In order to study the effect of adiabaticity on the generation of ferron, we have performed calculations with different switching on/off rates. We set $t_{\textrm{on}} = t_{\textrm{off}}-t_{\textrm{hold}} = t_{\textrm{switch}}$. 
The size of our simulation box is $48^3$ (in units of inverse $\kF$), which corresponds to the length of $37.70\xi$ along each dimension. 
The amplitude of the potential was fixed at $A_0=2\eF$. The width of the spin polarizing potential is set to $\sigma = 4.71\xi$. In~\fig{fig_etot_adia} we show evolution of the total energy for four different switching rates.
It is interesting to note that irrespectively of the switching on/off rate the nodal surface and consequently a stable ferron is always generated.
The difference between various scenarios is reflected in the local polarization, which attains the largest value in the case corresponding to the most rapid
ferron creation. In this case also the energy transfer to the system is the largest and beside the ferron structure many other excitations are created, mostly phonons. They interfere with the ferron and are responsible for fluctuations of the nodal surface, visible as large fluctuation for the phase difference in panel c) of Fig.~\ref{fig_etot_adia}.

\section{Conclusions}

We have presented various properties of the ferron depending on external conditions which include: 
total spin-imbalance, temperature, the rate of ferron creation.
The results indicate that the size of ferron is related to the total spin-polarization.
The static ferron solution becomes unstable at finite T which is much smaller than the critical temperature for the normal-to-superfluid phase 
transition. It implies that the creation of ferron needs to be achieved in a more adiabatic way limiting the energy transfer
to the system.

\section{Acknowledgements}

This work was supported by the Polish National Science Center (NCN) under Contract No. UMO-2016/23/B/ST2/01789 (BT,PM) and Contract No. UMO-2017/26/E/ST3/00428 (PK,GW). We acknowledge PRACE for awarding us access to resource Piz Daint based in Switzerland at Swiss National Supercomputing Centre (CSCS), decision No. 2017174125 and No. 2018194657. We also acknowledge the Global Scientific Information and Computing Center, Tokyo Institute of Technology for resources at TSUBAME3.0 (Project ID:hp190063). The static calculations were supported in part by PL-Grid Infrastructure and in part by Interdisciplinary Centre for Mathematical and Computational Modelling (ICM) of Warsaw University (grant No. GA76-13).

\end{document}